\documentclass[10pt,conference]{IEEEtran}

\usepackage{graphicx}

\makeatletter
\@namedef{figure*}{\@dblfloat{figure}}
\@namedef{endfigure*}{\end@dblfloat}
\def\subsection{\@startsection{subsection}{2}{\z@}{1.5ex plus 1.5ex minus 0.5ex}%
{0.7ex plus .5ex minus 0ex}{\normalfont\normalsize\itshape}}%
\makeatother

\begin{document}

\title{\LARGE A New SISO Algorithm with Application to Turbo Equalization$^1$}

\author{
\authorblockN{	Marcin Sikora and Daniel J. Costello, Jr.}
\authorblockA{	Department of Electrical Engineering, University of Notre Dame\\
		275 Fitzpatrick Hall, Notre Dame, IN 46556\\
		Email: {\tt \{msikora|dcostel1\}@nd.edu}
		\normalsize
		\\[8pt] Submitted to the ISIT 2005 on December 13th, 2004
	}
}

\maketitle

\footnotetext[1]{	This work was supported in part by NASA Grant NAG5-12792,
			NSF Grant CCR02-05310, Army grant DAAD16-02-C-0057,
			and the State of Indiana 21st Century Science and Technology fund.}

\begin{abstract}
In this paper we propose a new soft-input soft-output equalization algorithm,
offering very good performance/complexity tradeoffs.
It follows the structure of the BCJR algorithm,
but dynamically constructs a simplified trellis
during the forward recursion.
In each trellis section, only the $M$ states with the strongest forward metric
are preserved, similar to the M-BCJR algorithm.
Unlike the M-BCJR, however, the remaining states are not deleted, but rather merged
into the surviving states.
The new algorithm compares favorably with the reduced-state BCJR algorithm,
offering better performance and more flexibility, particularly for
systems with higher order modulations.
\end{abstract}

\section{Introduction}

\begin{figure}
   \centering
   \includegraphics[scale=0.70]{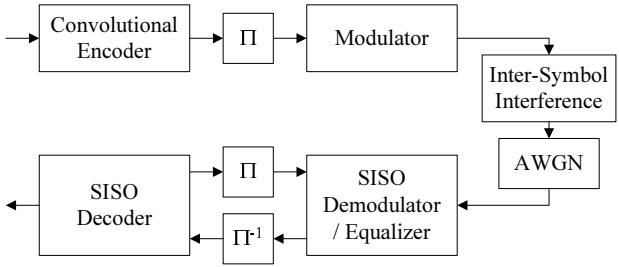}
   \caption{Communication system with turbo equalization.}
   \label{fig:systemmodel1}
\end{figure}

Efficient communication over channels introducing inter-symbol interference (ISI)
often requires the receiver to perform channel equalization.
Turbo equalization \cite{turbo_equal}
is a technique in which decoding and equalization are performed iteratively,
similar to turbo-decoding of serially-concatenated convolutional codes \cite{turbo_sccc}.
As depicted in Figure \ref{fig:systemmodel1},
the key element of the receiver employing this method is a soft-input soft-output (SISO)
demodulator/equalizer (from now on referred to as just an equalizer),
accepting a priori likelihoods of coded bits from the
SISO decoder, and producing their a posteriori likelihoods based on the noisy received signal.

The SISO algorithm that computes the exact values of the a posteriori
likelihoods is the BCJR algorithm \cite{bcjr}. The complexity of a BCJR equalizer
is proportional to the number of states in the trellis representing
the modulation alphabet and the ISI, and thus it is exponential
in both the length of the channel impluse response (CIR)
and in the number of bits per symbol in the modulator.
This can be a serious drawback in some scenarios, e.g.,
transmission at a high data rate over a radio channel,
where a large signal bandwidth translates to a long CIR,
and a high spectral efficiency translates to a large modulation alphabet.
Needed in such cases are alternative SISO equalizers with the ability to
achieve large complexity savings at a cost of small performance degradation.

There have been two main trends in the design of such SISOs. The first one relies on
reducing the effective length of the channel impulse response, either by
linear processing (see, e.g., \cite{tuechler_le}),
or interference cancellation via decision feedback.
A particularly good algorithm is this category is the reduced-state BCJR (RS-BCJR) \cite{rs_bcjr},
which performs the cancellation of the final channel taps on a per-survivor basis.
Iterative decoding with RS-BCJR
is very stable, thanks to the high quality of the soft outputs, but the receiver cannot use
the signal power contained in the cancelled part of the CIR.
Another trend is to adapt ``hard-output'' sequential algorithms \cite{sequential}
to produce soft outputs \cite{seq_revive}.
Examples in this category are the M-BCJR and T-BCJR algorithms \cite{m_bcjr},
based on the M- and T-algorithms, and the LISS algorithm \cite{liss} based on list sequential decoding.
These algorithms have no problem using the signal energy from the whole CIR,
and offer much more flexibility in choosing the desired complexity.
However, their reliance on ignoring unpromising paths in the trellis or tree
causes a bias in the soft output (there are more explored paths
with one value of a particular input
bit than another), which negatively affects the convergence of iterative decoding.

In this paper we present a new SISO equalization algorithm,
inspired by both the M-BCJR and RS-BCJR, which shares
many of their advantages, but few of their weaknesses.
We call this algorithm the M$^*$-BCJR algorithm,
since it resembles the M-BCJR in preserving only a fixed
number of trellis states with the largest forward metric.
Instead of deleting the excess states, however, the M$^*$-BCJR dynamically merges them with
the surviving states --- a process that shares some similarity to the static
state merging done on a per-survivor basis by the RS-BCJR.
For the sake of simpler notation, we present the operation of all BCJR-based algorithms,
including the M$^*$-BCJR, in the probability domain.
Each of them, however, can be implemented in the log domain
for better numerical stability.

The rest of the paper is structured as follows.
Section 2 describes the communication system
and the task of the SISO equalizer and introduces the notation.
Section 3 reviews the structure of the BCJR, M-BCJR, and RS-BCJR algorithms,
helping us to introduce the M$^*$-BCJR in Section 4.
Section 5 presents simulation results, and conclusions are given in Section 6.

\section{Communication system}

A communication system with turbo equalization is depicted in Figure \ref{fig:systemmodel1}.
The information bits are first arranged into blocks and encoded with a convolutional code.
The blocks of coded bits are permuted using an interleaver and mapped
onto a sequence of complex symbols by the modulator.
(In general, the modulator can have memory, but for simplicity we will
assume a memoryless mapper.) The channel acts as a discrete-time finite impulse response (FIR)
filter introducing ISI, the output of which is further corrupted by additive white Gaussian noise (AWGN).
We assume the receiver knows the ISI channel coefficients and the noise variance, and it attempts to recover
the information bits by iteratively performing SISO equalization and decoding.

\begin{figure}
   \centering
   \includegraphics[scale=0.70]{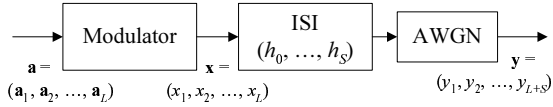}
   \caption{Part of the system to be ``soft-inverted'' by the SISO equalizer.}
   \label{fig:systemmodel2}
\end{figure}

The part of the system significant from the point of view of the equalizer is
shown in Figure \ref{fig:systemmodel2}.
Let ${\bf a}=({\bf a}_1,{\bf a}_2,...,{\bf a}_L)$ denote a sequence of $LK$ bits
entering the modulator, arranged into $L$ groups ${\bf a}_i=(a_i^1,a_i^2,...,a_i^K)$
of $K$ bits.
Each $K$-tuple ${\bf a}_i$ selects a complex-valued output symbol $x_i$
from a constellation of size $2^K$ to be transmitted.
The sequence of symbols ${\bf y}=(y_1,~y_2, ...,~y_{L+S})$ obtained at the receiver is modeled as
\begin{equation} \label{eq:channel1}
y_i = \sum_{j=0}^{S}h_jx_{i-j} + n_i,
\end{equation}
where $S$ is the memory of the channel,
$h_j$, $j=0,1,...,S$, are the channel coefficients,
and $n_i$, $i=1,2,...,L+S$, are i.i.d.~zero-mean complex-valued Gaussian random variables
with variance $\sigma^2$ per complex dimension.
Equation (\ref{eq:channel1}) assumes that $x_i$ is zero outside $i=1,2,...,L$.

The SISO equalizer for the above channel takes the received symbols ${\bf y}$ and the
a priori log-likelihood ratios $L_a(a_i^k)$ for each bit $a_i^k$, defined as
\begin{equation} \label{eq:apriori1}
L_a(a_i^k) = \log \frac{P(a_i^k=+1)}{P(a_i^k=-1)},
\end{equation}
and outputs the a posteriori L-values $L(a_i^k)$
\begin{equation} \label{eq:apost1}
L(a_i^k) = \log \frac{P(a_i^k=+1 | {\bf y})}{P(a_i^k=-1 | {\bf y})}.
\end{equation}
The values actually fed to the SISO decoder are extrinsic L-values, computed as
$L_e(a_i^k)=L(a_i^k)-L_a(a_i^k)$.

Let $\Lambda({\bf a})$ denote the joint probability that ${\bf a}$ was transmitted
and ${\bf y}$ was received.
Then (\ref{eq:apost1}) can be expressed as
\begin{equation} \label{eq:apost_lambda}
L(a^k_i) = \log \frac	{\sum_{{\bf a} : a^k_i=+1} \Lambda({\bf a})}
			{\sum_{{\bf a} : a^k_i=-1} \Lambda({\bf a})},
\end{equation}
where the summations are performed over all ${\bf a}$ consistent with $a^k_i=\pm1$.
Furthermore,
\begin{equation} \label{eq:lambda1}
\Lambda({\bf a}) = P({\bf a})
	\prod_{i=1}^{L+S} \frac{1}{\sqrt{2\pi\sigma^2}}
	    \exp(-\frac{1}{\sigma^2} ||r_i-\sum_{j=0}^{S}h_jx_{i-j}||^2),
\end{equation}
where $h_j$, $j=0,1,...,S$, and $\sigma^2$ are assumed known at the receiver
and $P({\bf a})$ is obtained from $L_a$ as
\begin{equation} \label{eq:apriori2}
P({\bf a}) = \prod_{i=1}^{L} \prod_{k=1}^{K} P(a_i^k),
\end{equation}
with
\begin{equation} \label{eq:apriori3}
P(a_i^k=\pm 1) = \frac {\exp(\pm L_a(a_i^k))}{1+\exp(\pm L_a(a_i^k))}.
\end{equation}

Since the number of paths involved in the summations of (\ref{eq:apost_lambda})
is extrememly large for realistic values of $K$ and $L$,
a practical algorithm seeks to simplify or approximate this calcualtion.

\section{SISO equalization}

\subsection{The BCJR algorithm}

The classical algorithm for efficiently computing (\ref{eq:apost_lambda}) by exploiting the trellis
structure of the set of all paths is the BCJR algorithm \cite{bcjr}.
By defining the state $s_i$ at time i as the past $S$ input symbol $K$-tuples ${\bf a}_i$,
$s_i=({\bf a}_{i-1},...,{\bf a}_{i-S})$, and a branch metric $\gamma(s_i,{\bf a}_i)$ as
\begin{equation} \label{eq:gamma}
\gamma(s_i,{\bf a}_i) = P({\bf a}_i) \frac{1}{\sqrt{2\pi\sigma^2}}
	    \exp(-\frac{1}{\sigma^2} ||r_i-\sum_{j=0}^{S}h_jx_{i-j}||^2),
\end{equation}
the path metric can be factored into
\begin{equation} \label{eq:lambda2}
\Lambda({\bf a}) = \prod_{i=1}^{L+S} \gamma(s_i,{\bf a}_i).
\end{equation}
For indices outside the range $i=1,...,L$, the variables ${\bf a}_i$ are regarded as
empty sequences $\phi$ with $P({\bf a}_i=\phi)=1$.

For every trellis branch $b_i=(s_i,{\bf a}_i,s_{i+1})$ starting in state $s_i$,
labeled by input bits ${\bf a}_i$, and ending in state $s_{i+1}$,
the BCJR algorithm computes the sum of the path metrics $\Lambda({\bf a})$
over all paths passing through this branch as
\begin{equation} \label{eq:bcjr_sum1}
\sum_{{\bf a}:b_i} \Lambda({\bf a}) = \alpha(s_i) \gamma(s_i,{\bf a}_i) \beta(s_{i+1}).
\end{equation}
The computation of the forward state metrics $\alpha(s_i)$ is performed in the
\emph{forward recursion} for $i=1,2,...,L+S-1$:
\begin{equation} \label{eq:bcjr_forward}
\alpha(s_{i+1})=\sum_{b_i=(s_i,{\bf a}_i,s_{i+1})} \alpha({s_i}) \gamma(s_i,{\bf a}_i),
\end{equation}
with the initial state value $\alpha(s_1)=1$.
Similarly, the \emph{backward recursion} computes the backward state metrics $\beta(s_i)$
for $i=L+S,L+S-1,...,2$:
\begin{equation} \label{eq:bcjr_backward}
\beta(s_{i})=\sum_{b_i=(s_i,{\bf a}_i,s_{i+1})} \gamma(s_i,{\bf a}_i) \beta({s_{i+1}}),
\end{equation}
with the terminal state value $\beta(s_{L+S+1})=1$.
With all $\alpha$'s, $\beta$'s, and $\gamma$'s computed, the summations over paths in
(\ref{eq:apost_lambda}) can be replaced by the summations over branches,
\begin{equation} \label{eq:bcjr_complete}
L(a^k_i) = \log \frac	{\sum_{b_i : a^k_i=+1} \alpha(s_i) \gamma(s_i,{\bf a}_i) \beta(s_{i+1})}
			{\sum_{b_i : a^k_i=-1} \alpha(s_i) \gamma(s_i,{\bf a}_i) \beta(s_{i+1})}.
\end{equation}
The \emph{completion} phase, in which (\ref{eq:bcjr_complete}) is evaluated for every $a_i^k$,
concludes the algorithm.

The complexity of the BCJR equalizer is proportional to the number of trellis states, $2^{KS}$.
The following subsections describe the operation of
the RS-BCJR \cite{rs_bcjr} and M-BCJR \cite{m_bcjr} algorithms, which preserve the
general structure of the BCJR, but instead operate on dynamically built simplified trellises
with a number of states controlled via a parameter.
In the original form of both algorithms, the construction of this simplified trellis occurs
during the forward recursion and is based on the values of the forward state metrics,
while the backward recursion and the completion phase just reuse the same trellis.

\subsection{The RS-BCJR algorithm}

The way we will describe the operation of the RS-BCJR algorithm is slightly different
from the presentation in \cite{rs_bcjr}, but is in fact equivalent.

Let us consider two states in the trellis,
\begin{eqnarray} \label{eq:rs_bcjr_1}
s_i=({\bf a}_{i-1}, ..., {\bf a}_{i-S'}, {\bf a}_{i-S'-1}, ..., {\bf a}_{i-S}), \\
s'_i=({\bf a}_{i-1}, ..., {\bf a}_{i-S'}, {\bf a}'_{i-S'-1}, ..., {\bf a}'_{i-S}),
\end{eqnarray}
differing only in the last $S-S'$ binary $K$-tuples.
Furthermore, consider two partial paths beginning in states $s_i$ and $s'_i$
and corresponding to the same partial input sequence ${\bf a}_{[i,L]}=({\bf a}_i, ..., {\bf a}_L)$.
Both paths are guaranteed to merge after $S-S'$ time indices,
and hence their partial path metrics are
\begin{eqnarray} \label{eq:rs_bcjr_2}
\lambda(s_i, {\bf a}_{[i,L]}) = \prod_{j=i}^{i+S-S'-1} \gamma(s_j, {\bf a}_j)
			\prod_{j=i+S-S'}^{L} \gamma(s_j, {\bf a}_j), \\
\lambda(s'_i, {\bf a}_{[i,L]}) = \prod_{j=i}^{i+S-S'-1} \gamma(s'_j, {\bf a}_j)
			\prod_{j=i+S-S'}^{L} \gamma(s_j, {\bf a}_j).
\end{eqnarray}
Additionally, close examination of (\ref{eq:gamma}) reveals that the difference between
$\gamma(s_j, {\bf a}_j)$ and $\gamma(s'_j, {\bf a}_j)$ for $j=i,...,i+S-S'-1$ is
not large. Hence, the difference between $\lambda(s_i, {\bf a})$ and $\lambda(s'_i, {\bf a})$,
for ${\bf a}_{[i,L]}$, is also not large.

The RS-BCJR equalizer relies on the above observation and, for some predefined $S'$,
declares states differing only in the last $S-S'$ binary $K$-tuples indistinguishable.
Every such set of states is subsequently reduced to a single state, by selecting
the state with the highest forward metric and \emph{merging} all remaining states into it.
Here, we define \emph{merging} of the state $s'_i$ \emph{into} $s_i$ as
updating the forward metric $\alpha(s_i) := \alpha(s_i)+\alpha(s'_i)$,
redirecting all trellis branches ending at $s'_i$ into $s_i$, and
deleting $s'_i$ from the trellis.
This reduction is performed during the forward recursion, and the $\gamma$'s
for the paths that originate from removed states need never be computed.
The trellis that results has only $2^{KS'}$ states, compared to $2^{KS}$ in the original trellis.
The same trellis is then reused in the backward recursion and the completion stage.

The RS-BCJR equalizer is particularly effective when the final coefficients
of the ISI channel are small in magnitude.
Furthermore, the reduced-state trellis retains the same branch-to-state ratio (branch density)
and has the same number of branches with $a_i^k=+1$ and $a_i^k=-1$ for any $i$ and $k$
--- properties that ensure a high quality for the soft outputs and good convergence
of iterative decoding.
Unfortunately, the RS-BCJR algorithm cannot use the signal power in the final $S-S'$ channel taps,
effectively reducing the minimum Euclidean distance between paths.
Moreover, the number of surviving states can only be set to a power of $2^K$,
which could be a problem for large $K$
(e.g., for a system with 16QAM modulation, equalization using 16 states could result in poor performance,
while 256 states could exceed acceptable complexity).

\subsection{The M-BCJR algorithm}

The M-BCJR algorithm is based on the M-algorithm \cite{sequential},
originally designed for the problem of maximum likelihood sequence estimation.
The M-algorithm keeps track only of the $M$ most likely
paths at the same depth, throwing away any excess paths.
In the M-BCJR equalizer this idea is applied to the trellis states during the forward recursion.
At every level $i$, when all $\alpha(s_i)$ have been computed, the $M$ states with the largest
forward metrics are retained, and all remaining states are deleted from the trellis
(together with all the branches that lead to or depart from them). The same trellis
is then reused in the backward recursion and completion phase.

In \cite{m_bcjr} it was shown that the M-BCJR algorithm performs well when the state reduction ratio
$2^{KS}/M$ is not very large. Also, unlike the RS-BCJR algorithm,
it can use the power from all the channel taps.
For small $M$, however, the reduced trellis is very
sparse, \emph{i.e.}, the branch-to-state ratio is much smaller than in the full trellis
and there is often a disproportion between the number of branches labeled with
$a_i^k=+1$ and $a_i^k=-1$ for any $i$ and $k$.
These factors reduce the quality of the soft outputs
and the convergence performance and may require an alternative way
of computing the a posteriori likelihoods
(like the Bayesian estimation approach presented in \cite{narayanan_mbcjr}).
Finally, the M-BCJR algorithm requires performing a partial sort
(finding the $M$ largest elements out of $M2^K$) at every trellis section,
which increases the complexity per state.

\section{The M$^*$-BCJR algorithm}

In this section we demonstrate how the concept of \emph{state merging} present in the RS-BCJR equalizer
can be used to enhance the performance of the M-BCJR algorithm. We call the resulting
algortihm the M$^*$-BCJR algorithm.

During the forward recursion the M$^*$-BCJR algorithm retains a maximum of $M$ states for any time index $i$.
Unlike the M-BCJR algorithm, however, the excess states are not deleted, but merely merged into
some of the surviving states. This means that none of the branches seen so far are deleted
from the trellis, but they are just redirected into a more likely state. The forward recursion
of the algorithm can be described as follows:

\begin{enumerate}

\item Set $i:=1$. For the initial trellis state $s_1$, set $\alpha(s_1):=1$.
Also, fix the set of states surviving at depth 1 to be $S_1:={s_1}$.

\item Initialize the set of surviving states at depth $i+1$ to an empty set, $S_{i+1} = \phi$.

\item For every state $s_i$ in the set $S_i$,
and every branch $b=(s_i,{\bf a}_i,s_{i+1})$ originating from that state,
compute the metric $\gamma(s_i,{\bf a}_i)$, and add $s_{i+1}$ to the
set $S_{i+1}$.

\item For every state $s_{i+1}$ in $S_{i+1}$ compute the forward state metric
as a sum of $\alpha(s_i)\gamma(s_i,{\bf a}_i)$ over all branches
$b=(s_i,{\bf a}_i,s_{i+1})$ visited in step 3 that end in $s_{i+1}$.

\item If the number of states in $S_{i+1}$ is no more than $M$,
proceed to step 8. Otherwise continue with step 6.

\item Determine the $M$ states in $S_{i+1}$ with the largest value of the forward
state metric. Remove all remaining states from $S_{i+1}$ and put them
in a temporary set $S'_{i+1}$.

\item Go over all states $s'_{i+1}$ in the set $S'_{i+1}$ and perform the following tasks for each of them:
\begin{itemize}
\item[-] Find a state $s_{i+1}$ in $S_{i+1}$ that differs from $s'_{i+1}$ by the
least number of final $K$-tuples ${\bf a}_j$.
\item[-] Redirect all branches ending in $s'_{i+1}$ to $s_{i+1}$.
\item[-] Add $\alpha(s'_{i+1})$ to the metric $\alpha(s_{i+1})$.
\item[-] Delete $s'_{i+1}$ from the set $S'_{i+1}$.
\end{itemize}

\item Increment $i$ by 1. If $i \le L+S-1$, go to step 2. Otherwise the forward recursion is finished.

\end{enumerate}

\begin{figure}
   \centering
   \includegraphics[scale=1.1]{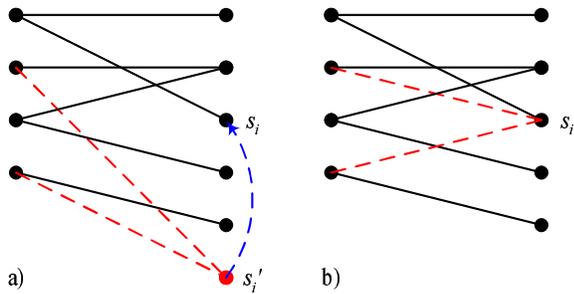}
   \caption{Trellis section a) before and b) after merging an excess state $s_i'$ into
   	 a surviving state $s_i$.}
   \label{fig:trellis2}
\end{figure}

The merging of $s_i'$ into $s_i$ in step 7 is also illustrated in Figure \ref{fig:trellis2}.
The backward recursion and the completion phase are subsequently performed
only over states remaining in the sets $S_i$ and only over visited branches (i.e., branches for which
the metrics $\gamma$ were calculated in step 3).

Just as for the M-BCJR, the M$^*$-BCJR algorithm can use the power from all channel taps
and offers full freedom in choosing the number of surviving states $M$.
At the same time, the M$^*$-BCJR never deletes visited branches, and hence it
retains the branch density of the full trellis and avoids a disproportion
between the number of branches labeled with $a_i^k=+1$ and $a_i^k=-1$.
As a result, the soft outputs generated by the M$^*$-BCJR equalizer ensure
good convergence of the iterative receiver.
Complexity-wise, the algorithm requires some additional processing per state (due to step 7)
and some additional memory per branch (the ending state must be remembered for each branch).
However, if we regard the calculation of the branch metrics $\gamma$ as the dominant
operation, the complexities of the M-BCJR, RS-BCJR, and M$^*$-BCJR equalizers are the same
for fixed $M=2^{KS'}$.

\section{Simulation results}

To evaluate the performance of the M$^*$-BCJR equalizer,
we considered two turbo-equalization systems.
Both systems used a recursive, memory 5, rate $1/2$ terminated convolutional code as an outer code.
The first system used BPSK modulation and a 5-tap channel (maximum 16 states),
and a block of 507 information bits (size 1024 DRP \cite{drp_interleaver} interleaver).
The second system used 16QAM modulation, but only a 3-tap channel (maximum 256 states),
and a block of 2043 information bits (size 4096 DRP interleaver).
The remaining parameters and the channel impulse responses are summarized in Table \ref{tbl:scenarios}.

\begin{figure*}[t]
   \begin{minipage}[c]{0.5\textwidth}
   	  \centering  a)\includegraphics[scale=0.55]{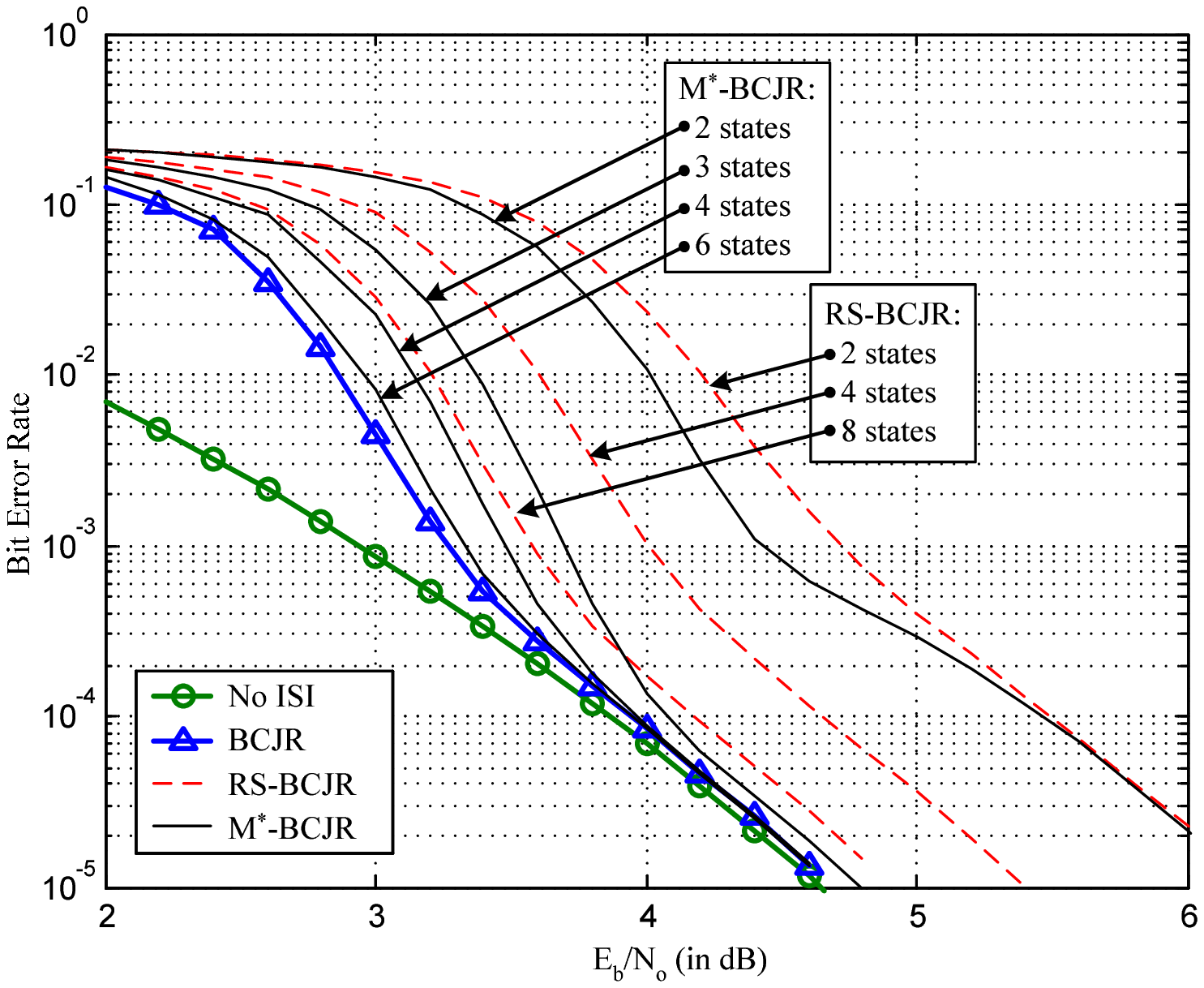}
   \end{minipage} %
   \begin{minipage}[c]{0.5\textwidth}
   	  \centering  b)\includegraphics[scale=0.55]{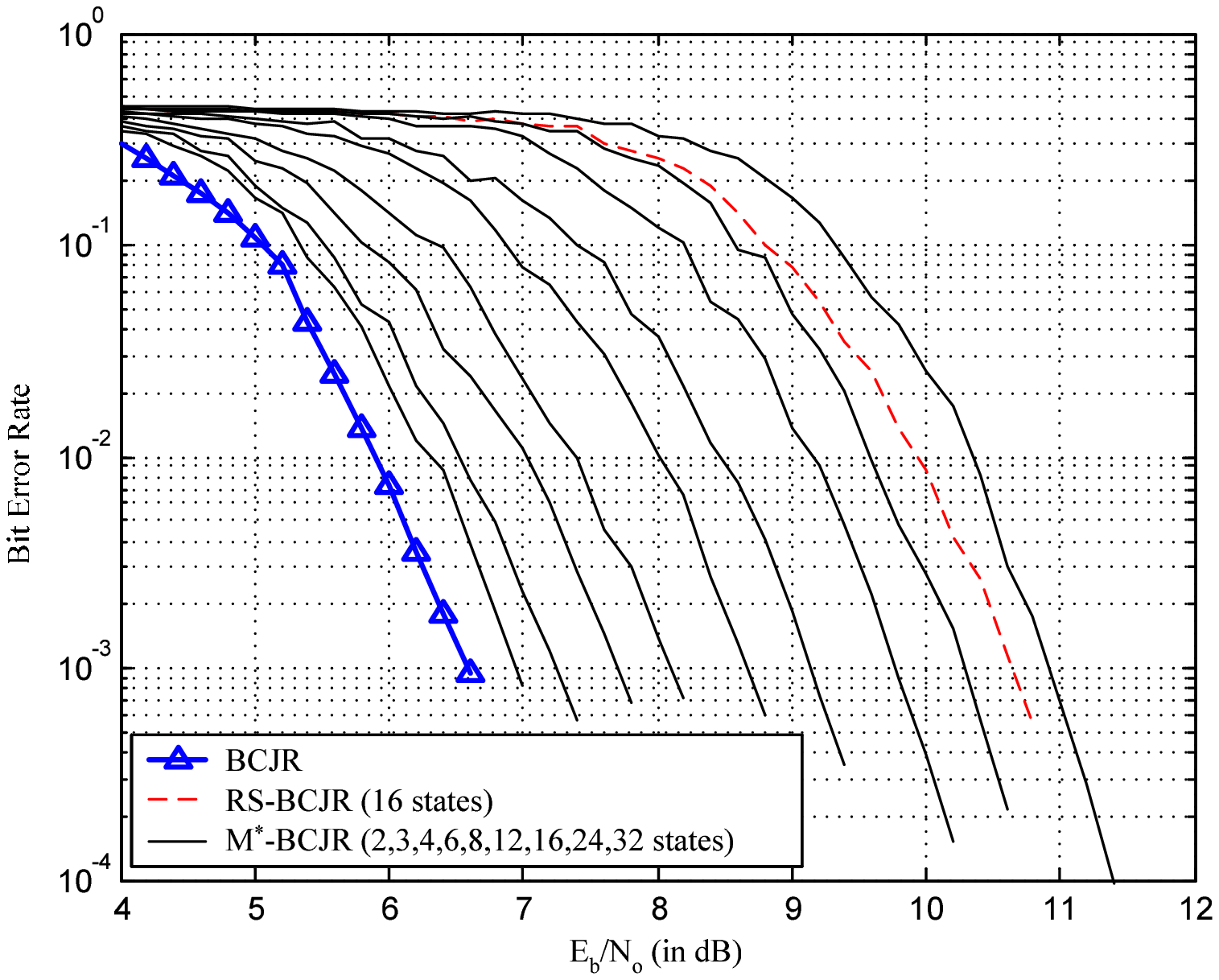}
   \end{minipage}
   \caption{Bit error rate of M$^*$-BCJR and RS-BCJR for a) scenario 1 (BPSK) and b) scenario 2 (16QAM).}
   \label{fig:bers}
\end{figure*}

\begin{figure*}[t]
   \begin{minipage}[c]{0.5\textwidth}
   	  \centering  a)\includegraphics[scale=0.55]{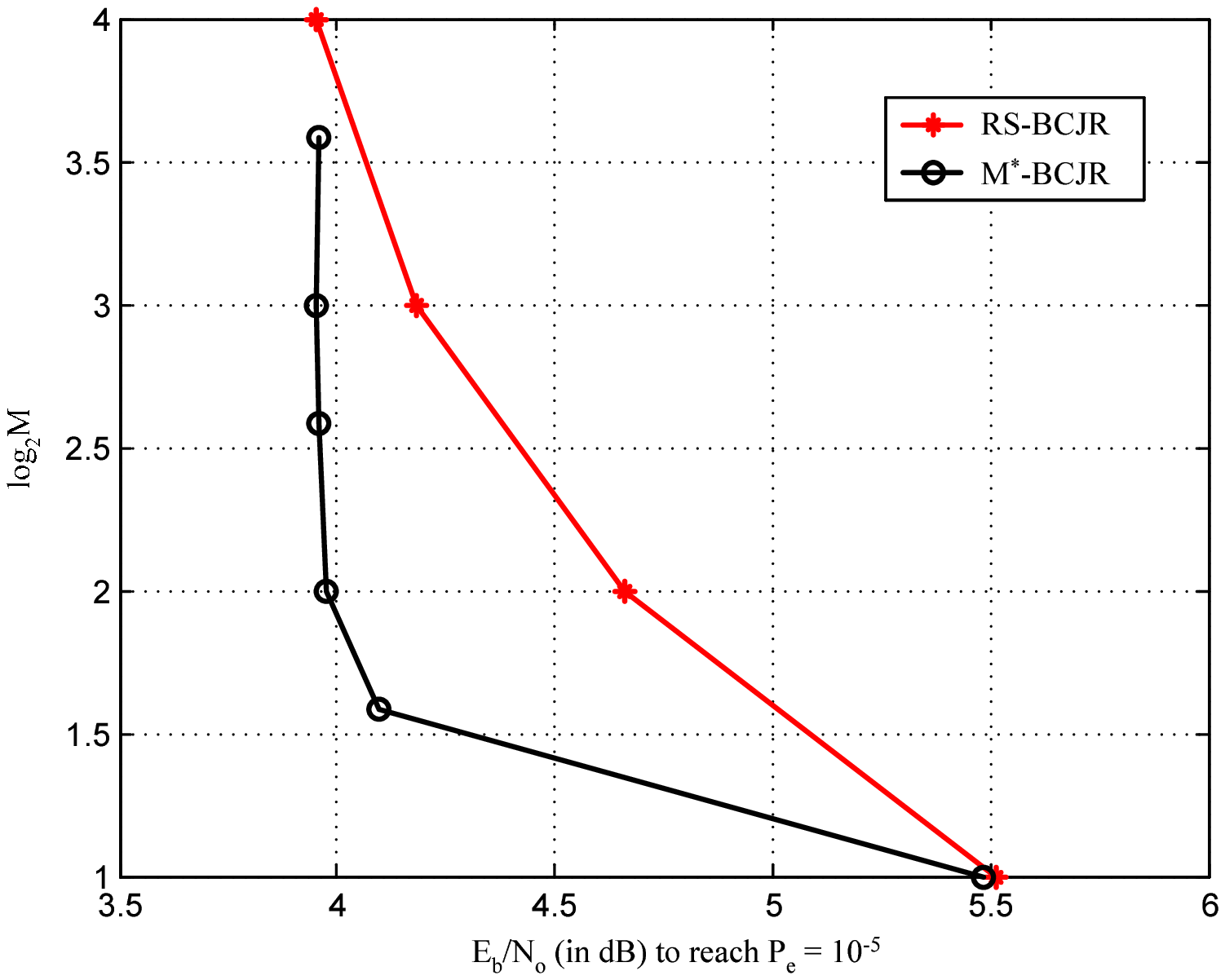}
   \end{minipage} %
   \begin{minipage}[c]{0.5\textwidth}
   	  \centering  b)\includegraphics[scale=0.55]{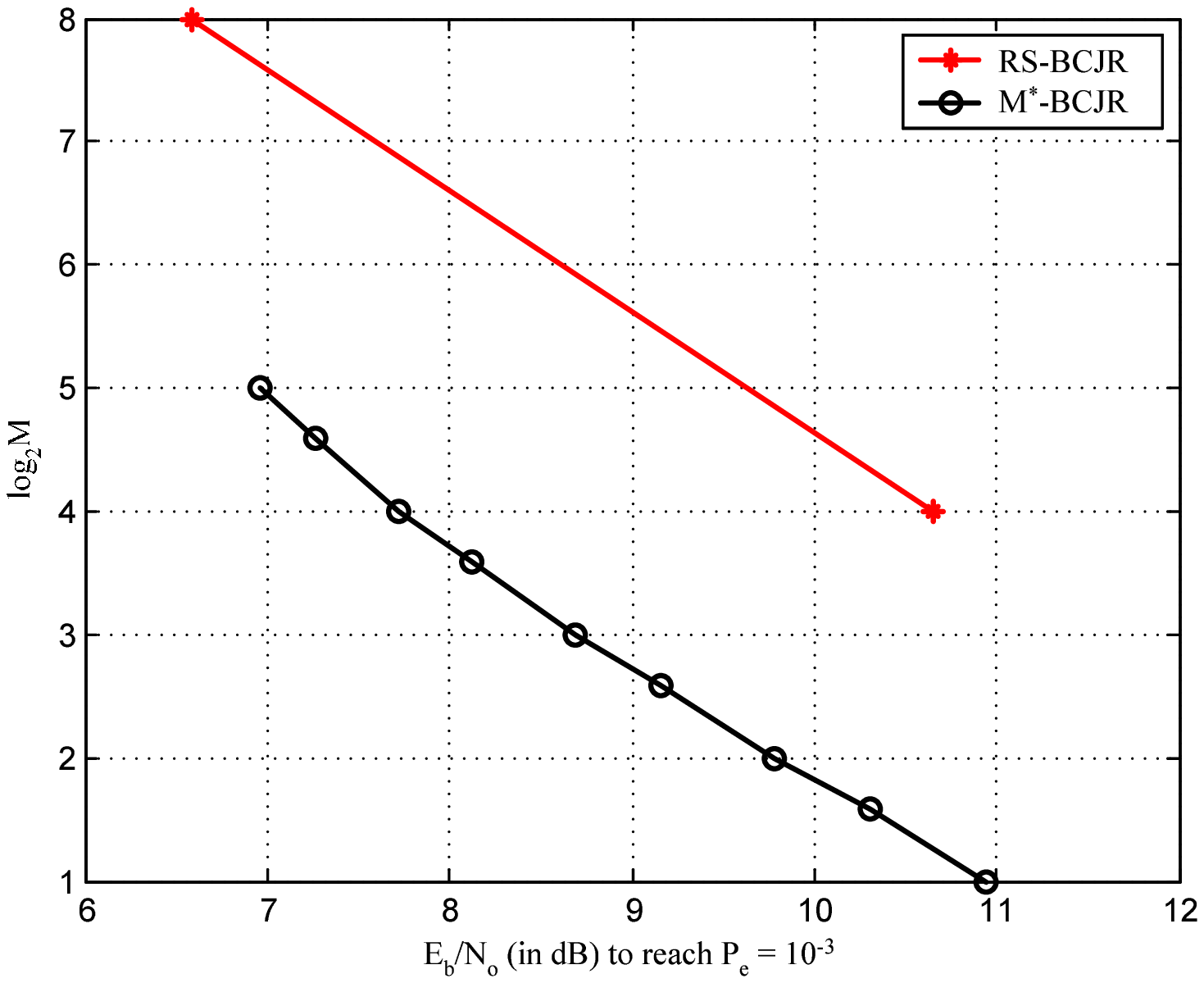}
   \end{minipage}
   \caption{Number of states vs. $E_b/N_o$ to reach the reference $P_e$ for
   		a) scenario 1 (BPSK, $P_e=10^{-4}$)
   		and b) scenario 2 (16QAM, $P_e=10^{-3}$).}
   \label{fig:states}
\end{figure*}

Both systems were simulated with the M$^*$-BCJR and RS-BCJR equalizers, for several values
of $M$ and $S'$. In each case we allowed the receiver to perform 6 iterations.
The bit error rates $P_e$ for a range of $E_b/N_o$ (average energy per bit over
noise spectral density) are plotted in Figure \ref{fig:bers}. To better
illustrate the complexity-performance tradeoffs achievable with both algorithms, we also
plotted the number of states $M$ or $2^{KS'}$ against the $E_b/N_o$ needed to achieve
certain $P_e$ ($10^{-4}$ for system 1 and $10^{-3}$ for system 2) in Figure
\ref{fig:states}.

The simulations demonstrate the superior performance of the M$^*$-BCJR equalizer.
In scenario 1, the M$^*$-BCJR equalizer with $3$ states outperforms
the RS-BCJR with $8$ states by 0.1 dB for $P_e$ below $10^{-4}$. When both algorithms use $4$ states,
the M$^*$-BCJR equalizer offers a 0.7 dB gain compared to the RS-BCJR.
In scenario 2, the M$^*$-BCJR with $16$ states achieves almost a 3 dB gain
over the RS-BCJR with the same number of states.

\begin{table}
   \renewcommand{\arraystretch}{1.2}
   \caption{Simulated turbo-equalization scenarios.}
   \label{tbl:scenarios}
   \begin{center}
   \begin{tabular}{|c|c|c|}
      \hline				&  Scenario 1 		&  Scenario 2 \\
      \hline	Outer code 		&  CC(2,1,5)		&  CC(2,1,5) \\
      \hline	Modulation		&  BPSK			&  16QAM \\
      \hline	Channel memory $S$     	&  4			&  2 \\
      \hline	CIR			&  $\{\sqrt{0.45},\sqrt{0.25},$			&  $\{1,1,1\}$ \\
      		$\{h_0,...,h_S\}$	&  $\sqrt{0.15},\sqrt{0.1},\sqrt{0.05}\}$	& \\
      \hline	BCJR states		&  16 			&  256 \\
      \hline	Interleaver size	&  1024 		&  4096 \\
      \hline	No. of iterations	&  6 			&  6 \\
      \hline
   \end{tabular}
   \end{center}
\end{table}

\section{Summary}

We have examined the problem of complexity reduciton in turbo equalization
for systems with large constellation sizes and/or long channel impulse responses.
We have defined the operation of merging one state into another and used
it to give an alternative interpretation of the RS-BCJR algorithm.
Finally we modified the M-BCJR algorithm, replacing the deletion of excess states by
the merging of these states into the surviving states.
The resulting algorithm, called the M$^*$-BCJR algorithm, was shown to
generate reduced-complexity trellises more suitable for SISO equalization
than those obtained by the RS-BCJR and M-BCJR algorithms.
Simulation results demonstrated very good performance for turbo-equalization
systems employing the M$^*$-BCJR, exceeding that of the RS-BCJR
even with much smaller complexities.

\end{document}